\documentclass[sigconf]{acmart}
\setcopyright{none}
\acmDOI{}
\acmISBN{}
\acmConference{}{}{}
\acmYear{}
\copyrightyear{}
\AtBeginDocument{%
  }

\setcopyright{none}
\acmDOI{}
\acmISBN{}

\usepackage{graphicx}
\usepackage{booktabs}
\usepackage{tabularx}
\usepackage[table]{xcolor}
\usepackage{colortbl}
\usepackage{float} 

\definecolor{TableHeader}{gray}{0.90}
\definecolor{TableHighlight}{gray}{0.94}
\begin{document}
\raggedbottom
\raggedbottom


\title[Spec Kit Agents: Context-Grounded Agentic Workflows]{\includegraphics[height=1.2em]{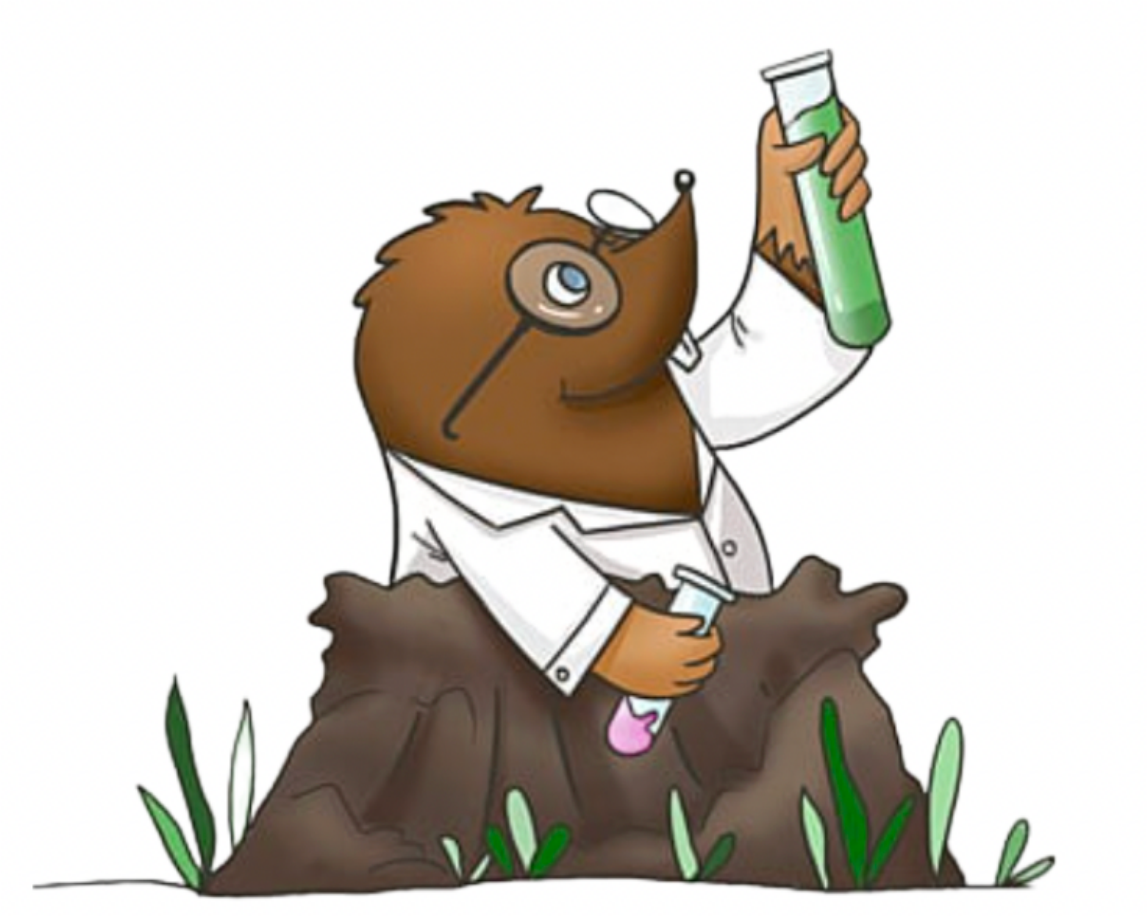}\texorpdfstring{Spec Kit Agents:\\Context-Grounded Agentic Workflows}{Spec Kit Agents: Context-Grounded Agentic Workflows}}

\author{Pardis Taghavi, Santosh Bhavani}


\begin{abstract}
Spec-driven development (SDD) with AI coding agents provides a structured workflow, but agents often remain ``context blind'' in large, evolving repositories, leading to hallucinated APIs and architectural violations. We present \textbf{Spec Kit Agents}
  a multi-agent SDD pipeline (with PM and developer roles) that adds phase-level, context-grounding hooks. Read-only probing hooks ground each stage (Specify, Plan, Tasks, Implement) in repository evidence, while validation hooks check intermediate artifacts against the environment. We evaluate 128 runs covering 32 features across five repositories. Context-grounding hooks improve judged quality by +0.15 on a 1–5 composite LLM-as-judge score. (+3.0\% of the full score; Wilcoxon signed-rank, $p<0.05$) while maintaining 99.7--100\% repository-level test compatibility. We further evaluate the framework on SWE-bench Lite, where augmentation hooks improve baseline by 1.7\%, achieving 58.2\% Pass@1.
\end{abstract}


\begin{teaserfigure}
  \centering
  \includegraphics[width=0.7\textwidth]{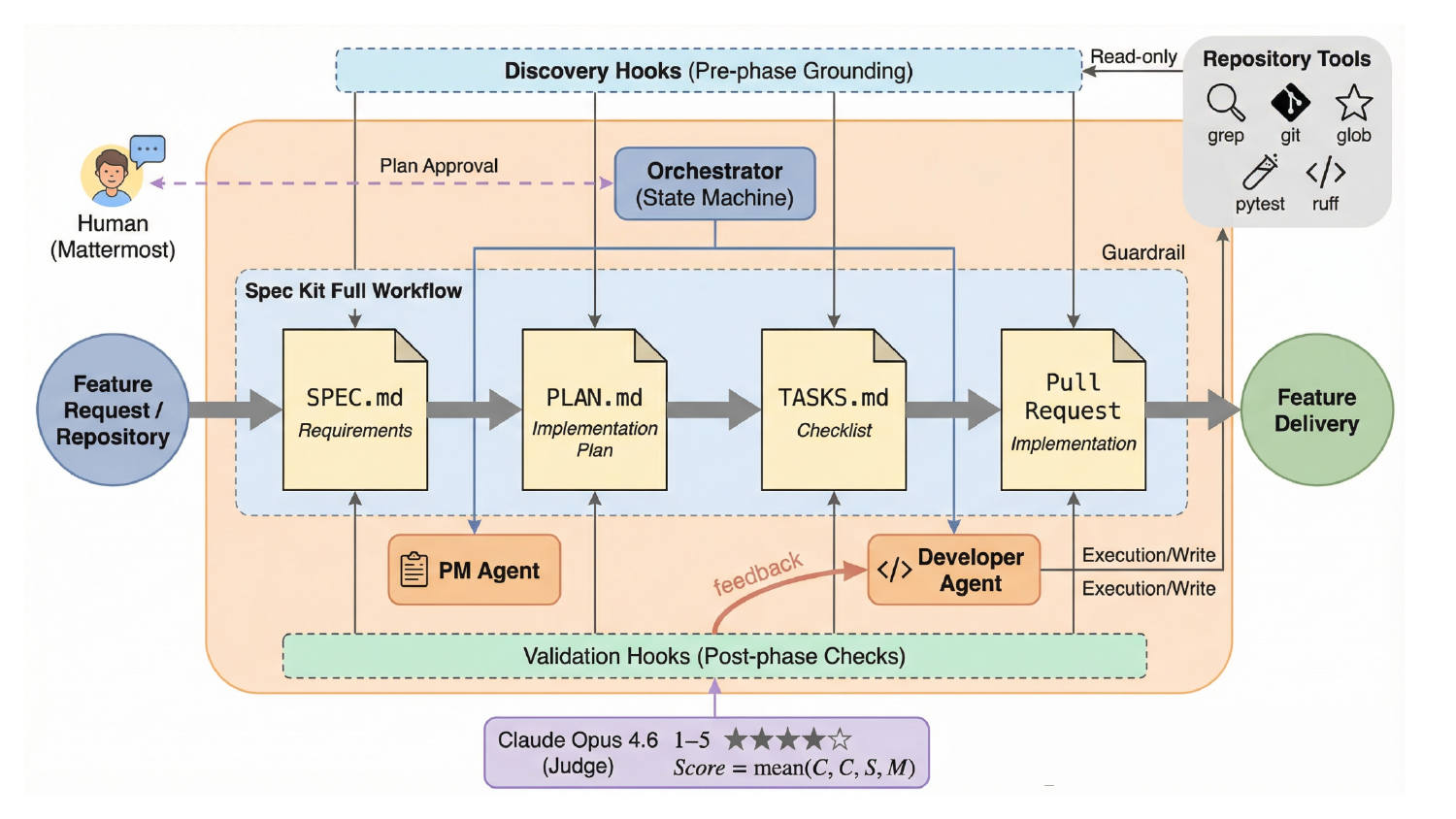}
  \caption{Overview of the Spec Kit Agents workflow.}
  \Description{E}
  \label{fig:teaser}
\end{teaserfigure}


\maketitle

\section{Introduction}
Large language models (LLMs) have made it practical to automate substantial portions of software development, but end-to-end feature delivery in real repositories remains brittle. Modern coding assistants are effective at local edits, yet multi-step tasks in evolving codebases frequently fail for reasons including missing context about the current architecture, stale assumptions about dependencies, and mismatches with repository conventions. These failures tend to compound across stages such as planning, task decomposition, and implementation leading to wasted iterations and unreliable outcomes.
Spec-driven development (SDD) is a promising response to this brittleness. Rather than asking an agent to generate code immediately, SDD externalizes intermediate artifacts (e.g., a specification, an implementation plan, and a task checklist) that make intent explicit and provide a structured audit trail. GitHub's Spec Kit~\cite{SpecKitDocs} operationalizes this idea as a staged workflow (Specify $\rightarrow$ Plan $\rightarrow$ Tasks $\rightarrow$ Implement), optionally gated by plan review. In principle, this ``reasoning before coding'' structure should improve reliability and debuggability.

In practice, however, structured workflows do not eliminate a core failure mode we refer to as \emph{context blindness}: the agent's intermediate artifacts can be internally coherent while being incompatible with the repository as it exists. Common symptoms include referencing non-existent APIs, proposing file paths that do not exist, and violating local architectural or stylistic conventions. When these errors are discovered late during implementation or test execution the agent often backtracks, revises earlier artifacts, or introduces additional inconsistencies. We present \textbf{Spec Kit Agents}, an orchestrated multi-agent SDD pipeline that addresses context blindness by making grounding and validation explicit workflow operations. Spec Kit Agents augments the Spec Kit stages with a \emph{context-grounding layer}: (i) \emph{discovery} hooks that perform read-only probing before each stage to collect repository evidence (relevant files, conventions, dependencies, history), and (ii) \emph{validation} hooks that check intermediate artifacts and, after implementation, execute project checks (e.g., tests and linters) when applicable. This design keeps grounding and validation outside the core agent prompts, enabling auditable traces and selective tool access.
\paragraph{Contributions.}
\begin{itemize}
  \item \textbf{System.} Spec Kit Agents, a multi-agent SDD pipeline (state-machine orchestrator + PM and developer roles) with a context-grounding layer that runs pre-phase discovery and post-phase validation hooks.
  \item \textbf{Context-grounding design.} A phase scoped grounding and validation interface that operates over explicit artifacts (SPEC/ PLAN/ TASKS), enabling transparent auditing and least privilege tool access.
  \item \textbf{Evaluation.} An empirical study over $128$ experimental runs covering 32 unique feature tasks across 5 repositories, reporting judged quality, latency, and repository-level test compatibility. We also report controlled comparisons of Baseline, Augmented, Full, and Full-Augmented configurations, together with Discovery-only and Validation-only ablations, and evaluate generalization on SWE-bench Lite.
\end{itemize}

Across 128 task instances (32 unique feature tasks across 5 open-source repositories), Spec Kit Agents yields a consistent improvement in judged quality (+0.15 on a 1–5 composite LLM-as-judge score) while maintaining high test pass rates (99.7--100\%). These gains come with additional overhead in the full workflow family due to extra phases and context-grounding execution; accordingly, we interpret latency within each budget family rather than across families. Overall, the primary benefit is not a dramatic jump in average score, but earlier detection and prevention of compounding context errors in multi-step agentic workflows.

\section{Related Work}

\subsection{Multi-agent orchestration and agentic workflows.} Recent LLM-agent research has moved from single-model prompting to \emph{agentic workflows} that decompose tasks into structured stages and often assign specialized roles across agents \cite{wang2023llmagentsurvey, guo2024large}. Common workflow primitives include closed-loop reasoning with tool use (e.g., ReAct) \cite{yao2023react} and search over intermediate reasoning states to improve planning and execution \cite{yao2023tree}. Multi-agent frameworks such as AutoGen, CAMEL, and MetaGPT, along with newer orchestration systems, emphasize role specialization, coordination, and interaction protocols \cite{wu2023autogen,li2023camel,hong2023metagpt,chen2023agentverse,agentorchestra2025,evolvingorchestration2025,marco2024}. Benchmarks likewise show that orchestration design materially affects agent performance across tasks \cite{liu2023agentbench,agentboard2024}.
Where prior work primarily emphasizes orchestration and collaboration among agents, our work targets workflow reliability under context limitations. Spec Kit Agents adds a context-grounding layer that performs phase-level grounding and validation outside the core agent prompts.
\subsection{Tool-augmented grounding for agents.} A challenge in agentic systems is grounding decisions in external evidence rather than relying on parametric memory. Retrieval-augmented generation improves factuality and supports knowledge-intensive tasks by conditioning outputs on retrieved documents \cite{lewis2020rag,asai2023selfrag}, while browser- and tool-augmented systems show that allowing agents to query external sources and cite evidence can improve task success \cite{nakano2021webgpt,press2022compositionalitygap}. More broadly, tool use has been studied through modular routing to external tools or experts (e.g., MRKL-style systems) \cite{karpas2022mrkl} and through learned tool-use behaviors acquired via training or self-supervision \cite{schick2023toolformer,qin2023toolllm,patil2023gorilla,li2023apibank}. In software engineering, agents extend these ideas to repository-level grounding through file search, code navigation, executable actions, and exploration~\cite{yang2024sweagent,zhang2024autocoderover,wang2024openhands,xia2024agentless}.
Most prior approaches treat grounding as an in-trajectory behavior of the same agent that plans and generates, making it sensitive to prompt design and context-window noise. In contrast, Spec Kit Agents makes grounding an explicit workflow primitive: read-only discovery hooks probe repository state before each phase, and validation hooks check intermediate artifacts against executable signals. This shifts grounding from best-effort retrieval to phase-scoped evidence collection, making it more repeatable, inspectable, and less coupled to the main agent’s generation.
\subsection{Verification, context-grounding, and tool-based validation}
Reliability work on LLM agents includes self-critique and iterative refinement methods that use feedback to improve later attempts \cite{shinn2023reflexion,madaan2023selfrefine,gou2023critic,chen2025revisitselfdebug,yuan2025reveal}, as well as rule-based constraint approaches that steer behavior through explicit principles \cite{bai2022constitutionalai,runtimeenforce2025}. Many agent pipelines also rely on tool-based validation signals such as tests, linters, and structured checks, especially in repository-level tasks where executable feedback provides a strong correctness signal \cite{jimenez2023swebench,yang2024sweagent}. Benchmarks further show that verification and feedback design materially affect end-to-end reliability \cite{liu2023agentbench}.
Our contribution differs in both \emph{when} and \emph{what} we validate. Rather than concentrating verification after implementation, we validate intermediate artifacts (SPEC/PLAN/TASKS) before code generation, catching hallucinated APIs, invalid paths, and architectural mismatches early while retaining post-implementation executable checks as a final gate. More broadly, we treat tool-based validation not as a single end stage filter, but as repeated phase-specific context-grounding hooks that reduce compounding errors across agentic workflows.

\section{Method}
We present \textbf{Spec Kit Agents}, including its orchestration logic, tool interfaces, and context-grounding mechanisms. We also describe the execution and evaluation protocol used in our experiments.

\subsection{System Overview and Workflow}
Spec Kit Agents is a multi-agent system for feature delivery in existing repositories. The system consists of (i) an \emph{orchestrator} implemented as a state machine, (ii) a \emph{product manager (PM) agent} responsible for clarifying requirements and prioritization, and (iii) a \emph{developer agent} responsible for producing intermediate artifacts and implementing code changes. Agents communicate through a centralized messaging platform, which also supports human intervention at defined checkpoints (e.g., plan approval).
The developer agent follows the Spec Kit workflow to generate intermediate artifacts and then implement the feature. In the \emph{Full} workflow variants, the developer agent produces three intermediate artifacts before implementation: \texttt{SPEC.md} (requirements and acceptance criteria), \texttt{PLAN.md} (an implementation plan with file-level touchpoints), and \texttt{TASKS.md} (an executable checklist). The implementation stage then executes the plan and opens a pull request in the target repository. In \emph{Baseline} variants, the agent skips all intermediate artifacts and proceeds directly to implementation.

\subsection{Context-Grounded Agentic Workflows Layer}
We introduce a context-grounding layer that provides phase-scoped grounding and validation for the developer agent. The context-grounding hooks are invoked at workflow boundaries and operate over explicit artifacts (e.g., \texttt{SPEC.md}, \texttt{PLAN.md}, and \texttt{TASKS.md}) rather than being embedded inside the developer’s main prompt.

\textbf{Discovery hooks (pre-phase grounding).} Before each phase, a read-only prober gathers evidence about the codebase using repository inspection tools (e.g., globbing, grep, and \texttt{git} history). The goal is to surface project-specific conventions, existing APIs, and relevant modules so that subsequent generation is conditioned on concrete, localized context rather than generic priors. For example, for persistence-related features, discovery can identify existing logging formats or storage abstractions and steer the agent away from introducing unsupported dependencies.

\textbf{Validation hooks (post-phase checks).} After each phase, a validator checks the generated artifact for internal consistency and repository compatibility. For earlier artifacts, validation focuses on structural and referential constraints (e.g., whether file paths referenced in \texttt{PLAN.md} exist, whether required libraries are present, and whether the task list is feasible and properly ordered). After implementation, validation executes repository checks (e.g., unit tests and linters) to detect regressions. This design front-loads error detection by catching hallucinated paths, missing dependencies, or infeasible plans before code generation compounds mistakes.

\textbf{Tool access control.} The context-grounding hooks validate each phase by probing the codebase before and after reasoning steps, ensuring specifications are grounded in existing project conventions and plans are verified against installed dependencies. The PM agent is restricted to repository analysis and version-control inspection. The developer agent is permitted to edit files and run repository commands required to implement features. Discovery hooks are read-only, while validation hooks extend discovery permissions with execution privileges for project checks (e.g., \texttt{pytest}, \texttt{ruff}, and JavaScript test runners) when applicable.

\subsection{Models, Tools, and Execution Environment}
Spec Kit Agents separates \emph{generation} from \emph{evaluation}. The agentic workflow (PM and developer agents) is executed through Claude Code CLI, routed to an Anthropic-compatible endpoint backed by MiniMax-M2.5. Using a single execution interface ensures consistent tool invocation, logging, and run control across all experiments.
Quality is evaluated independently using Claude Opus 4.6 as an LLM-as-judge. Outputs are scored on a 1--5 scale along four dimensions: \emph{completeness}, \emph{correctness}, \emph{style}, and \emph{maintainability}, and the composite score is their mean. This separation reduces self-evaluation bias by isolating scoring from the agent's prompts and tool access. We additionally conduct a small blinded human review on a subset of outputs using the same rubric.
We log prompts, tool calls, intermediate artifacts, and execution traces for each run. Rate-limited runs are excluded from latency analyses but retained for quality reporting when a pull request artifact is available; completion rates include such runs.

\subsection{Experimental Protocol and Configurations}
\textbf{Configurations.} We evaluate four primary configurations: (i) \emph{Baseline}, which skips intermediate artifacts and proceeds directly to implementation; (ii) \emph{Augmented}, which follows the same direct-to-implementation flow and adds discovery and validation hooks; (iii) \emph{Full}, which executes the full Spec Kit workflow; and (iv) \emph{Full-Augmented}, which adds discovery and validation hooks to Full. To isolate context-grounding effects, we also evaluate \emph{Discovery-only} (pre-phase hooks only) and \emph{Validation-only} (post-phase hooks only) ablations.
\textbf{Budgets and timeouts.} Each phase is subject to bounded timeouts to control end-to-end runtime. Human-facing checkpoints for plan-review are auto approved. End-to-end, Baseline and Augmented runs use a 40-minute budget, while Full and Full-Augmented runs use a 90-minute budget. Runs exceeding these limits are terminated and marked as failures.
\textbf{Success criteria.} A run is considered successful if it produces a pull request in the target repository, includes at least one file modification, and completes without critical execution errors (e.g., authentication failures or tool-permission violations). Quality is assessed post hoc using the judge model; in analysis, composite scores below 3.0 are treated as requiring manual review.
\section{Experiments}

\subsection{Evaluation Setup}
We evaluate \textbf{Spec Kit Agents} on 32 feature tasks across five repositories: FastAPI, Airflow, Dexter, Plausible Analytics, and Strapi. Each task is run under four configurations: \emph{Baseline}, \emph{Augmented}, \emph{Full}, and \emph{Full-Augmented}. The task set spans multiple change types, including API additions, configuration changes, new modules, refactors, and test updates; Appendix~A lists a subset of tasks and categories.
FastAPI and Airflow are Python repositories evaluated with \texttt{pytest -q}; Dexter and Strapi are TypeScript repositories; Plausible Analytics is primarily Elixir with supporting JavaScript. For each task, the agent receives a natural-language feature request and executes the assigned workflow end-to-end, producing a pull request when successful.
Our primary outcome is judged quality, measured by an independent LLM-as-judge (Claude Opus 4.6) using a 1--5 composite score. We also report wall-clock completion time, test-suite compatibility based on post-change repository test execution, and failure category for unsuccessful runs. Generalization to SWE-bench Lite is evaluated separately. For statistical comparisons, we treat each feature task as a paired subject across conditions and use the Wilcoxon signed-rank test for paired analyses of judged quality and wall-clock completion time.

\subsection{Quality Results}
Table~\ref{tab:quality-merged} reports judged quality, with the overall score computed as a feature-count-weighted average across repositories. 
In the 40-minute workflow family (\emph{Baseline}, \emph{Augmented}), the developer agent skips intermediate artifacts (\texttt{SPEC.md}/\texttt{PLAN.md}/\texttt{TASKS.md}) and proceeds directly to implementation; in the 90-minute family (\emph{Full}, \emph{Full-Augmented}), those artifacts are produced before coding.
Within the 90-minute workflow family, \emph{Full-Augmented} achieves the strongest overall quality, improving from 3.51 to 3.66 (+0.15) relative to \emph{Full}. On the paired subset of completed tasks, this difference is statistically significant (Wilcoxon signed-rank, $p<0.05$). Gains appear across repositories, with especially strong improvements on FastAPI and Plausible. To complement the LLM-based evaluation, we also conduct a blinded human preference study on paired tasks completed successfully under both \emph{Full} and \emph{Full-Augmented}. Evaluators compare anonymized pull requests shown in random order and may select either version or a tie. Table~\ref{tab:human-preference} summarizes the resulting pairwise judgments. Repository-level test-suite compatibility remains high across configurations, indicating that the quality gains do not come at the expense of breaking existing project behavior.

\begin{table}[t]
  \centering
  \caption{Quality scores. Shaded cells indicate the best result within each workflow family.}
  \label{tab:quality-merged}
  \footnotesize
  \setlength{\tabcolsep}{5pt}
  \begin{tabular}{lcccccc}
    \toprule
    \textbf{Condition} & \textbf{Over.} & \textbf{F-API} & \textbf{Airf.} & \textbf{Dext.} & \textbf{Plau.} & \textbf{Strap.} \\
    \midrule
    Baseline & 3.46 & 3.21 & \cellcolor{TableHighlight}{3.75} & \cellcolor{TableHighlight}{3.65} & 3.30 & 3.25 \\
    Augmented & \cellcolor{TableHighlight}{3.50} & \cellcolor{TableHighlight}{3.58} & 3.56 & 3.31 & \cellcolor{TableHighlight}{3.45} & \cellcolor{TableHighlight}{3.55} \\
    \midrule
    Full & 3.51 & 3.10 & 3.35 & 3.90 & 3.48 & 3.61 \\
    Full-Augmented & \cellcolor{TableHighlight}{3.66} & \cellcolor{TableHighlight}{3.52} & \cellcolor{TableHighlight}{3.44} & \cellcolor{TableHighlight}{4.00} & \cellcolor{TableHighlight}{3.64} & \cellcolor{TableHighlight}{3.69} \\
    \bottomrule
  \end{tabular}
\end{table}

\begin{table}[t]
\footnotesize
\centering
\caption{Blinded human preference on paired pull-request comparisons.}
\label{tab:human-preference}
\begin{tabular}{lrrrrr}
\toprule
\textbf{Comp.} & \textbf{Tasks} & \textbf{Votes} & \textbf{Full} & \textbf{Tie} & \textbf{Full-Aug.} \\
\midrule
Full vs.\ Full-Aug. & 6 & 60 & 19 & 33 & 8 \\
\bottomrule
\end{tabular}
\end{table}

\subsection{Ablation Results}
We ablate context-grounding components by enabling only pre-phase discovery or only post-phase validation. Table~\ref{tab:ablation} reports the resulting quality and runtime relative to the \emph{Full} baseline (3.51). Both partial variants improve over \emph{Full}, with \emph{Validation-only} outperforming \emph{Discovery-only}. The combined design achieves the strongest result, suggesting that the two components are complementary.
\subsection{Latency Results}
We report wall-clock completion time on completed runs only. \emph{Baseline} and \emph{Augmented} use a 40-minute budget, whereas \emph{Full} and \emph{Full-Augmented} use a 90-minute budget, latency is compared only within each budget family. Context-grounding hooks add only modest overhead in the 40-minute family, but a larger cost in the 90-minute family due to the longer workflow and repeated hook execution. We therefore view latency as a quality--runtime trade-off.
\begin{table}[t]
  \centering
  \caption{Ablation of phase-level context-grounding components relative to the \emph{Full} baseline (3.51).}
  \label{tab:ablation}
  \footnotesize
  \setlength{\tabcolsep}{4pt}
  \begin{tabular}{lcccc}
    \toprule
    \textbf{Condition} & \textbf{Qual.} & \textbf{$\Delta$\%} & \textbf{Time} & \textbf{Description} \\
    \midrule
    Discovery-only  & 3.53 & +0.57\% & 25.5 min & Pre-phase grounding \\
    Validation-only & 3.57 & +1.71\% & 31.2 min & Post-phase checks \\
    Full-Augmented  & \textbf{3.66} & \textbf{+4.27\%} & 37.2 min & Both hooks enabled \\
    \bottomrule
  \end{tabular}
\end{table}

\begin{table}[t]
\footnotesize
\centering
\caption{Within-family latency comparisons (completed runs only).}
\label{tab:latency-family}
\begin{tabular}{lrrrr}
\toprule
\textbf{Comparison} & \textbf{A (min)} & \textbf{B (min)} & \textbf{$\Delta$ (min)} & \textbf{$n_{\text{pairs}}$} \\
\midrule
Baseline vs.\ Augmented  & 14.4 & 15.5 & +1.1  & 15 \\
Full vs.\ Full-Augmented & 24.0 & 37.2 & +13.2 & 16 \\
\bottomrule
\end{tabular}
\end{table}

\subsection{SWE-bench Lite Results}   
To assess generalization beyond our custom repository tasks, we evaluate Spec Kit Agents on SWE-bench Lite, a standard benchmark of 300 real-world software engineering issues. Table~\ref{tab:swebench-lite} compares our framework against prior SOTA. Spec Kit Agents achieves a \textbf{56.5\%} pass rate in the baseline configuration and \textbf{58.2\%} with context-grounding hooks enabled. All experiments in this paper use MiniMax-M2.5 as the base model, however, the proposed orchestration framework is model-agnostic and readily generalizes to other API-accessible models.Additional implementation and failure-mode details are provided in Appendices B and C.

                                                                               
  \begin{table}[t]                                                                                                                                               
    \small
    \centering                                                                                                                                                   
    \caption{Comparative analysis on SWE-bench Lite. Spec Kit Agents shows competitive performance using MiniMax-M2.5.}
    \label{tab:swebench-lite}                                                                                                                                    
    \begin{tabularx}{\linewidth}{@{}lXr@{}}                                                                                                                      
    \toprule                                                                                                                                                     
    \rowcolor{TableHeader}
    \textbf{Framework} & \textbf{Primary LLM} & \textbf{Pass@1} \\
    \midrule
    Aider~\cite{aider2024swebenchlite} & GPT-4o \& Claude 3 Opus & 26.33 \\
    Moatless Tools~\cite{moatlessTools2024} & Claude 3.5 Sonnet & 38.00 \\
    OpenHands~\cite{wang2024openhands} & CodeAct v2.1 & 41.67 \\
    DARS Agent~\cite{aggarwal2025dars} & Claude 3.5 Sonnet + DeepSeek R1 & 47.00 \\
    
    SWE-Agent~\cite{yang2024sweagent} & Claude 4 Sonnet & 56.67 \\

    \midrule
    \rowcolor{TableHighlight}
    \textbf{Spec Kit Agents (Ours, Baseline)} & \textbf{MiniMax-M2.5} & \textbf{56.5} \\
    \rowcolor{TableHighlight}
    \textbf{Spec Kit Agents (Ours, Augmented)} & \textbf{MiniMax-M2.5} & \textbf{58.2} \\
    \bottomrule
    \end{tabularx}
    \end{table}

\section{Conclusion}
We presented \textbf{Spec Kit Agents}, a multi-agent, spec-driven development workflow that augments Spec Kit with phase-scoped discovery and validation context-grounding hooks. Across 128 runs covering 32 features, the context-grounded full workflow achieves the strongest overall quality, indicating that explicit repository-grounded orchestration improves reliability. The gains are consistent and stem from stronger alignment between the specification, the discovered repository context, and the final implementation. This improvement comes with additional runtime overhead, making the approach most appropriate for higher-risk or high complexity tasks. Overall, the results support explicit context-grounded orchestration as a practical design principle for more dependable autonomous software engineering.

\bibliographystyle{ACM-Reference-Format}
\bibliography{sample-base}

\AtBeginEnvironment{thebibliography}{\sloppy\emergencystretch=10em\hyphenpenalty=50\exhyphenpenalty=50}

\appendix

\section{Representative Task Set}
Table~\ref{tab:feature-list} provides a representative subset of the task set used in the custom repository evaluation. These examples illustrate the diversity of repositories and change types considered in the study.

\begin{table*}[t]
\footnotesize
\centering
\caption{Illustrative subset of tasks used in the custom repository evaluation.}
\label{tab:feature-list}
\begin{tabular}{llll}
\toprule
\textbf{Repository} & \textbf{Task ID} & \textbf{Category} & \textbf{Description} \\
\midrule
Dexter  & \texttt{dex-01}  & \texttt{config\_change} & Add \texttt{--json} flag for JSON output \\
Dexter  & \texttt{dex-02}  & \texttt{new\_module}    & Session persistence with the \texttt{--session} flag \\
Dexter  & \texttt{dex-03}  & \texttt{api\_endpoint}  & Telegram bot integration \\
Dexter  & \texttt{dex-04}  & \texttt{new\_module}    & Streaming response mode \\
Dexter  & \texttt{dex-05}  & \texttt{refactor}       & Portfolio analysis module \\
\midrule
FastAPI & \texttt{fapi-01} & \texttt{new\_module}    & SSE streaming support \\
FastAPI & \texttt{fapi-02} & \texttt{refactor}       & Validation error improvements \\
FastAPI & \texttt{fapi-03} & \texttt{new\_module}    & Plugin system \\
FastAPI & \texttt{fapi-04} & \texttt{api\_endpoint}  & OpenAPI schema enhancements \\
FastAPI & \texttt{fapi-05} & \texttt{new\_module}    & Typed middleware \\
\midrule
Airflow & \texttt{af-01}   & \texttt{new\_module}    & Error message improvements \\
Airflow & \texttt{af-02}   & \texttt{test}           & DAG testing utilities \\
Airflow & \texttt{af-03}   & \texttt{new\_module}    & Custom metrics support \\
Airflow & \texttt{af-04}   & \texttt{config\_change} & Type annotations \\
Airflow & \texttt{af-05}   & \texttt{new\_module}    & Memory monitoring \\
\midrule
Plausible & \texttt{pla-01} & \texttt{new\_module}   & Funnel visualization with conversion tracking \\
Plausible & \texttt{pla-03} & \texttt{new\_module}   & Advanced filter builder (AND/OR conditions) \\
Plausible & \texttt{pla-05} & \texttt{api\_endpoint} & GraphQL API for analytics data \\
\midrule
Strapi   & \texttt{str-01} & \texttt{new\_module}   & Content version history with restore \\
Strapi   & \texttt{str-03} & \texttt{new\_module}   & Redis query result caching \\
Strapi   & \texttt{str-07} & \texttt{new\_module}   & Algolia search plugin \\
\bottomrule
\end{tabular}
\end{table*}

\section{Reproducibility Details}
\subsection{Repository-Level Analysis on SWE-bench Lite}
The gain from augmentation is not uniform across SWE-bench repository families. We observe stronger improvements when failures are tightly coupled to unit-tested, test-adjacent code paths (e.g., pytest- or linter-facing fixes), where discovery and validation hooks can directly align the implementation with executable checks. By contrast, augmentation is less reliable on repositories such as django and matplotlib, where many failures originate in deeper application/library logic (ORM state transitions or visualization-state interactions) that are only weakly exposed by local unit tests. In SymPy-like cases, mathematically subtle edge conditions can also be under-specified by the available tests, so context-grounding hooks may anchor on incomplete signals. Overall, augmentation helps most when tests directly exercise the underlying defect, and helps less when fixes require integration context, database state, or net-new functionality beyond the tested path.

\subsection{Model and Tool Versions}
Table~\ref{tab:model-versions} summarizes the primary models and tools used in the experiments. We separate generation, execution, and evaluation roles to make the pipeline explicit.

\begin{table}[t]
\footnotesize
\centering
\caption{Model and tool versions used in the experiments.}
\label{tab:model-versions}
\begin{tabularx}{\columnwidth}{@{}l l l X@{}}
\toprule
\textbf{Role} & \textbf{System} & \textbf{Version} & \textbf{Notes} \\
\midrule
Generator         & MiniMax-M2.5    & N/A      & Primary LLM used for Spec Kit Agents execution \\
Execution wrapper & Claude Code CLI & 2.1.50   & Invoked via \texttt{claude -p} \\
Judge evaluator   & Claude Opus 4.6 & 20250501 & LLM-as-judge model ID \\
\bottomrule
\end{tabularx}
\end{table}

\subsection{Prompting and Artifacts}
The context-grounding layer uses structured prompts for pre-phase discovery and post-phase validation. In \emph{Full} and \emph{Full-Augmented}, hooks run at \texttt{specify}, \texttt{plan}, \texttt{tasks}, and \texttt{implement}, and prompts are parameterized by the current workflow state and intermediate artifacts (\texttt{SPEC.md}, \texttt{PLAN.md}, \texttt{TASKS.md}). In \emph{Baseline} and \emph{Augmented}, no intermediate artifacts are generated; execution proceeds directly to implementation (with implementation-stage hooks when enabled). Prompt templates are part of the experimental pipeline and are parameterized by workflow stage and intermediate artifacts (e.g., \texttt{SPEC.md}, \texttt{PLAN.md}, \texttt{TASKS.md}).

\subsection{Configuration Files}
The following configuration files were used to support reproducibility:
\begin{itemize}
  \item \texttt{config.yaml}: system-level configuration (timeouts, tool permissions, and workflow settings),
  \item \texttt{experiments/features.yaml}: task definitions for the reported feature set,
  \item \texttt{experiment\_runner.py}: execution and orchestration logic, and
  \item \texttt{quality\_evaluator.py}: LLM-as-judge scoring implementation.
\end{itemize}

\subsection{Execution Environment}
Experiments were conducted on a MacBook Pro (Apple Silicon) with 32\,GB RAM running macOS. The execution pipeline relies on remote model inference; typical network latency to the serving endpoint was approximately 200--500\,ms. As described in the main text, transient rate limits were handled via exponential backoff, and rate-limit events were logged in the run metadata.

\section{Failure Taxonomy}
To make unsuccessful runs more interpretable, we categorize failures by their primary cause. These categories align with the success criteria used in the main text.

\begin{itemize}
  \item \textbf{Budget timeout:} the run exceeds the end-to-end time budget and is terminated.
  \item \textbf{Human-checkpoint timeout:} an approval or clarification step is not resolved within the configured timeout.
  \item \textbf{Artifact validation failure:} an intermediate artifact fails phase-level validation (e.g., invalid paths, missing dependencies, or infeasible task ordering).
  \item \textbf{Execution or environment failure:} the run encounters an authentication issue, tool-permission violation, or another execution-layer failure.
  \item \textbf{Repository-check failure:} implementation completes, but post-change tests or linters fail.
  \item \textbf{Incomplete implementation:} no pull request is produced, or no meaningful file modification is made.
  \item \textbf{Rate-limited or interrupted run:} progress is interrupted by API throttling or another transient execution failure before completion.
\end{itemize}

\end{document}